\documentclass[aps,prl,a4paper,twocolumn,superscriptaddress,floatfix,showpacs]{revtex4}

\usepackage{graphicx}
\usepackage{dcolumn}
\usepackage{amsmath}
\usepackage{bm}

\begin{document}

\title{Disorder Potentials near Lithographically Fabricated
Atom Chips}
\author{P.~Kr\"uger}
\email{krueger@physi.uni-heidelberg.de}
\homepage{http://www.atomchip.net}
\author{L.~M.~Andersson}
\author{S.~Wildermuth}
\author{S.~Hofferberth}
\author{E.~Haller}
\author{S.~Aigner}
\author{S.~Groth}
\affiliation{Physikalisches Institut, Universit\"at Heidelberg,
69120 Heidelberg, Germany}
\author{I.~Bar-Joseph}
\affiliation{Department of Condensed Matter Physics, The Weizmann
Institute of Science, Rehovot 76100, Israel}
\author{J.~Schmiedmayer}
\affiliation{Physikalisches Institut, Universit\"at Heidelberg,
69120 Heidelberg, Germany}

\date{September 24, 2004}

\begin{abstract}
We show that previously observed large disorder potentials in
magnetic microtraps for neutral atoms are reduced by about two
orders of magnitude when using atom chips with lithographically
fabricated high quality gold layers. Using one dimensional
Bose-Einstein condensates, we probe the remaining magnetic field
variations at surface distances down to a few microns.
Measurements on a 100 $\mu$m wide wire imply that residual
variations of the current flow result from {\em local} properties
of the wire.
\end{abstract}

\pacs{39.90.+d, 03.75.Be}

\maketitle

Trapping and manipulating cold neutral atoms in microtraps near
surfaces of atom chips is a promising approach towards a full
quantum control of matter waves on small scales \cite{Fol02}. In a
number of experiments a variety of trapping, guiding and
transporting potentials have been realized using current carrying
wires \cite{Rei99,Hae01a,Rei02,Mue99,Dek00,Fol00,Cas00}, atom
manipulation with electric fields was integrated on an atom chip
\cite{Kru03}, coherent dynamics of internal atomic hyperfine
states was observed \cite{Tre04} and easy formation of
Bose-Einstein condensates (BEC) was demonstrated
\cite{Ott01,Hae01b,Lea02,Sch03b}.

The full potential of atom chip experiments is only accessible if
the potentials can be miniaturized to a scale of typically 1
$\mu$m or below where appreciable tunnelling rates between
separated traps can be reached, and efficient atom-atom coupling
between atoms in neighboring trap sites \cite{Cal00a} can be
achieved. While the fabrication of structure sizes $< 1$ $\mu$m is
not problematic, unintended potential roughness has been reported
to severely alter the trapping at surface distances $d$ below
$\sim100$ $\mu$m, resulting in a longitudinal fragmentation of
elongated clouds \cite{For02,Lea03,Jon03,Est04}. Such disorder
potentials have been observed near macroscopic wires and atom
chips fabricated by electroplating techniques
\cite{Drn98,For02b,Lev03}.

Strongly confining trapping and guiding potentials on atom chips
are formed by the subtraction of two magnetic fields, the field of
a current carrying wire and a (homogeneous) bias field (side guide
configuration \cite{Den99a}). The remaining field at the potential
minimum is determined by the angle between wire field and bias
field. A small change of the current direction may result in a
significant change in the trapping potential.

The observed disorder potentials have been attributed to
inhomogeneous magnetic field components $\Delta B$ in the
direction {\em parallel} to the current carrying wire creating the
trapping field $B$ \cite{For02,Lea03,Jon03,Est04}. It has been
suggested that such field components could be derived from
fabrication inhomogeneities, surface roughness \cite{Kas03,Sch04}
and residual roughness of the wire borders \cite{Wan04}. The model
of Wang et al. \cite{Wan04} provides a full quantitative
explanation of the potentials found near electroplated gold wires
\cite{Est04}.

\begin{figure}
    \includegraphics[width=\columnwidth]{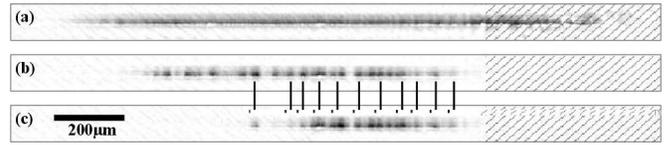}
    \caption{\label{fig:frag_image}{\em In situ} absorption images
    of atom clouds positioned at a chip surface distance of $\sim
    5$ $\mu$m above a current carrying wire (cross section
    $3.1\times 10$ $\mu$m$^2$). Parts of the images do not fully
    represent the atom density distribution since the imaging
    light beam was obstructed by bonding wires (hatched regions).
    a) Thermal atoms show no fragmentation. b)
    BECs display a much higher sensitivity and residual
    disorder potentials cause a fragmentation of the cloud.
    c) A longitudinal displacement of the BEC by tuning the
    trapping potential shows that the disorder potential is stable
    in position.}
\end{figure}

In this Letter, we report on a dramatic reduction of the disorder
potentials in our experiments. For cold thermal atoms ($T\sim 1$
$\mu$K) we do not observe fragmentation of the trapped clouds even
when they are brought to distances of 3 $\mu$m from the surface of
a current carrying wire. Yet, we are able to measure a small
residual potential roughness near the wire by creating BECs close
to the surface (Fig.\ \ref{fig:frag_image}).

\begin{figure}
    \includegraphics[width= \columnwidth]{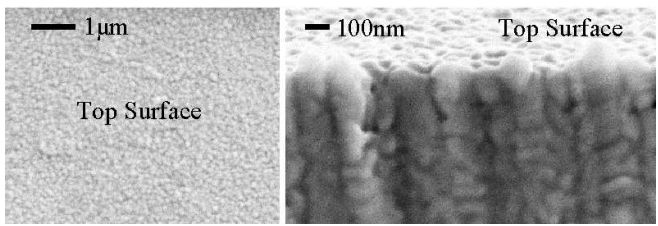}
    \caption{\label{fig:fab} Scanning electron microscope images
    of the chip wire surface (left) and edges (right). The grain sizes of
    $<100$ nm determine both the surface and edge roughness.}
\end{figure}

We attribute this reduction of the disorder potentials to our atom
chip fabrication method. We obtain chips with very smooth wire
structures by adapting a standard microchip fabrication process to
the production of our atom chips \cite{Gro04,foot}. Masks written
by electron beam lithography are used to structure a several
micron thick, high quality gold layer on a semiconductor wafer
using a lift-off procedure. The result is a smooth gold mirror
with precise gaps defining the current path in the wire
\cite{Fol00,Fol02,Gro04}. Fig.\ \ref{fig:fab} shows electron
microscope images of the gold surface and the wire edges.

In our experiment, more than $10^8$ $^{87}$Rb atoms are
accumulated a few mm from the chip surface which serves directly
as a mirror for a reflection magneto-optical trap (MOT)
\cite{Rei99}. These atoms are subsequently transferred to a purely
magnetic trap and cooled to $\sim 5$ $\mu$K by radio frequency
(RF) evaporation. Both the MOT and the magnetic trap are based on
copper wire structures mounted directly underneath the chip
\cite{Wil04}. The resulting sample of $\gtrsim 10^6$ atoms is then
loaded to the selected chip trap and location, where a second
stage of RF evaporative cooling creates either a BEC or thermal
cloud just above the critical condensation temperature.

\begin{figure}
    \includegraphics[angle=0,width= \columnwidth]{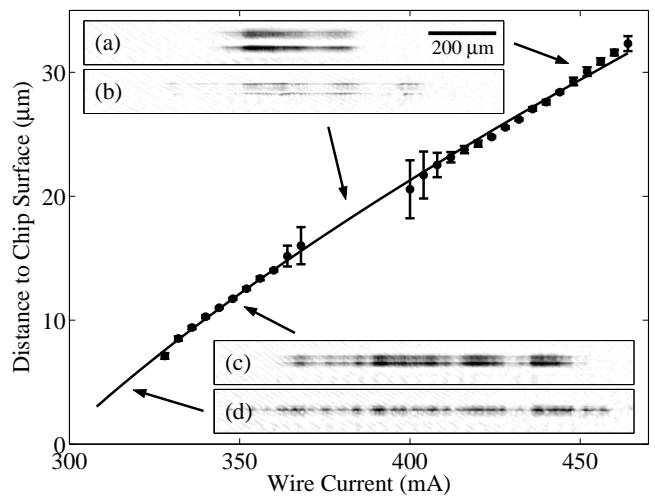}
    \caption{\label{fig:height} Distance $d$ of the BEC from the (mirror)
    chip surface as a function of wire current ($3.1\times 100$ $\mu$m$^2$ wire). Atoms
    near the surface produce a double image when illuminated by an
    inclined imaging beam as shown in the inserts a)-c). The
    imaging beam together with the chip surface produces a fringe
    pattern that makes distance measurements less reliable for
    certain surface distances (b). For clouds closer than
    $\sim 5$~$\mu$m from the surface, the two images merge (d). To
    determine $d$ also in these cases we use an extrapolation according to a best
    fit (solid line) with the exact bias field strength as only fitting
    parameter. The fitting model takes the finite size of the wire
    into account.}
\end{figure}

We image the atomic clouds near the surface {\em in situ} by
resonant absorption imaging with $\sim 3.5$ $\mu$m resolution. In
order to determine the cloud's distance from the surface $d$, we
slightly incline the imaging light with respect to the chip mirror
surface by $\sim 25$ mrad. For sufficiently small $d$ ($<100$
$\mu$m) this leads to a duplicated absorption image \cite{Sch03b}
(Fig.\ \ref{fig:height}).

While the wire currents are very well known, the strength of the
external bias fields has to be calibrated with measurements of $d$
at sufficiently large $d$. As our imaging resolution does not
allow to measure $d$ for very close surface approaches, we use the
calibrated values of the bias fields together with the measured
wire currents to infer $d$ in these cases.

We have probed the residual potential roughness for various
trapping geometries based on a 100 $\mu$m and several 10 $\mu$m
wide wires at atom-surface distances down to 3~$\mu$m. The global
parameters of the atomic cloud like atom number and temperature
are determined by the ballistic expansion of the cloud in
time-of-flight measurements.

With thermal atoms we always observe smooth longitudinal
absorption profiles inside the trap (Fig.\ \ref{fig:frag_image}a),
independent of the wire used to form the trap and the position of
the atomic cloud. For the closest approach of $d=3$ $\mu$m, a
cloud at $T=1$ $\mu$K remains un-fragmented within our detection
resolution, even when summing up many realizations of the
experiment to reduce measurement noise. Assuming that the atomic
density profile follows the Boltzmann distribution $n\sim
\exp(-V/k_BT)$, we can put an upper limit to the residual magnetic
field roughness $\Delta B/B<2\times 10^{-4}$ where $B$ is the
field produced by the wire at that distance.

BECs are a much more sensitive probe of potential roughness. The
relevant energy scale, given by the chemical potential $\mu$, can
be orders of magnitude smaller than the temperature of thermal
atoms. Figs.\ \ref{fig:frag_image}b and c show typical absorption
images of fragmented BECs at $d=5$~$\mu$m from one of the 10
$\mu$m wide wires. Altering the longitudinal confinement by
varying the current in an independent auxiliary wire leads only to
an overall displacement of the cloud while the local disorder
potential variations remain stable in their positions. Over many
months of experiments no change was observed in the position of
the fragments.

The inserts of Fig.\ \ref{fig:height} show absorption profiles of
BECs at various heights $d$ above the 100 $\mu$m wide wire. As the
surface is approached, the longitudinal trapping potential becomes
flatter, thus the BECs extend over a longer stretch of the wire.
As $d$ is increased, the strength of the disorder potentials is
reduced and the typical length scale of fragmentation increases.
For $d\gtrsim 30$ $\mu$m, virtually no fragmentation is detected,
even with a BEC.

\begin{figure}
    \includegraphics[width= \columnwidth]{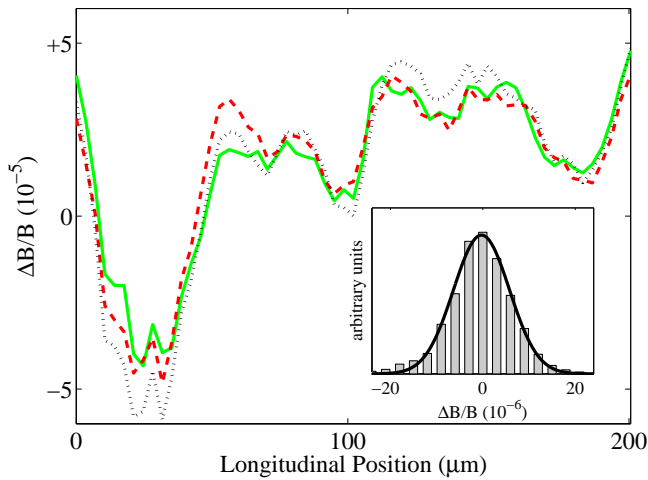}
    \caption{\label{fig:height_movie} (color online) Longitudinal potential
    profiles measured with BECs at a constant distance of $d=10$ $\mu$m
    from the surface of the 100 $\mu$m broad wire. The different
    traces were measured at different currents and are normalized
    to the respective trapping fields. The bias field (10~G, 20~G, 30~G;
    black dotted, solid green, dashed red lines, respectively)
    was adapted in order to keep $d$ constant. The
    insert shows a histogram of the deviations of the
    curves. The width of the distribution ($\sigma \sim 8\times10^{-6}$) is
    similar to the shot to shot variations of different realizations
    of the same experiment with equal wire currents.}
\end{figure}

For a quantitative analysis, we extract longitudinal density
profiles $n_\mathrm{1d}$ from the {\em in situ} absorption images
and calibrate them with the absolute atom number derived from
time-of-flight images taken under equal experimental conditions.
For $n_\mathrm{1d}\lesssim 100$ $\mu$m$^{-1}$ ($^{87}$Rb atoms),
the confinement is of one dimensional (1d) character, i.e.\ the
transverse single particle ground state energy exceeds the
chemical potential $\mu$ of the BEC. In our experiments this
condition is always fulfilled, and the actual potential
experienced by the atoms can be reconstructed according to $V(x) =
-2\hbar\omega_\perp a_\mathrm{scat} n_\mathrm{1d}(x)$ where
$a_\mathrm{scat}\approx 5.6$ nm is the $^{87}$Rb scattering
length. This expression is derived under the assumption of a
constant (global) $\mu$ in a 1d Thomas-Fermi (TF) approximation
\cite{Ber03}. This is strictly valid only in an equilibrium state
of the system. This may not be the case in our experiment over the
entire length of the BEC ($\sim 1$ mm). Similar to the
observations previously made in an optical dipole double well
potential \cite{Shi04}, a variation of $\mu$ on longitudinal
length scales $>200$ $\mu$m is maintained longer than the life
time of the BEC if strong potential barriers separate the
different fragments of the condensate. We have confirmed the
validity of the TF-approach for shorter wavelength components by
monitoring the $n_\mathrm{1d}(x)$ profile over the entire lifetime
of the condensate. We observe that the reconstructed potential
fluctuations at wavelengths $\lesssim 200$ $\mu$m remain constant.
We limit the further analysis to length scales shorter than 200
$\mu$m.

To assess whether the observed disorder potentials are magnetic in
origin we have varied the wire current while adapting the bias
field so that the BECs were trapped at fixed distances from the
wire. Magnetic disorder potentials stemming from an irregular
current flow should scale linearly with the current in the wire
$I$. Figure \ref{fig:height_movie} shows an example of relative
potential variations $\Delta B/B$ reconstructed from BECs
positioned at $d=10$ $\mu$m for three different currents. To
quantify the consistency between the measurements, we compare the
shot to shot variations of $\Delta B/B$ with equal currents to
those with different currents. We find equal widths of the
residual differences between the graphs. We conclude that within
the statistical similarity of the $\Delta B/B$ distributions
($\sim 3\times 10^{-6}$), we can exclude any current independent
sources of disorder potentials such as electrostatic patch effects
\cite{McG04} at the scale of $10^{-13}$ eV for $d>5$~$\mu$m.

In order to study the source of the irregular current flow we have
measured the variation of the disorder potentials with $d$. Wires
of two different widths, 10 $\mu$m and 100 $\mu$m, were used. The
main observation is that the scaling of the amplitude and the
frequency spectrum of the disorder potentials with $d$ for the two
wires are very similar. For $d < 50$ $\mu$m this would not be the
case if edge fluctuations were dominating as can be derived from
the edge fluctuation model \cite{Wan04}.

\begin{figure}
    \includegraphics[angle=-90, width= \columnwidth]{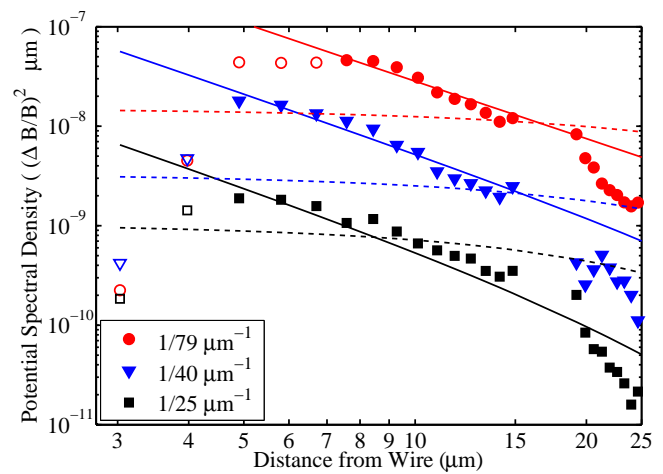}
    \caption{\label{fig:scaling} (color online) Spectral power density of the
    disorder potentials near the 100 $\mu$m wide wire for
    three spatial frequencies.
    The open symbols correspond to data where the detected signal
    is limited by the chemical potential $\mu$ of the BEC.  For
    these data points the complete depth of the potential cannot be
    measured, and they were omitted in the analysis.
    The solid lines are best fits according to a local
    fluctuating current path model, the dashed lines show best
    fits to the model outlined in \cite{Wan04}. For both models
    the only fitting parameter is the strength of current path
    fluctuation at the respective spatial frequency $k$.}
\end{figure}

For the 100 $\mu$m wide wire, Fig.\ \ref{fig:scaling} shows
potential spectral densities (PSD) of the disorder potential at
three different spatial frequencies $k$. In the examined
$d$-range, the potentials scale more strongly with $d$ than they
would for dominating edge fluctuations \cite{Wan04} for all
frequency components. We interpret the clear difference in slope
of the experimental data and the wire edge model as an indication
that {\em local} current path deviations are important. Such
deviations can occur due to inhomogeneous conductivity or top
surface roughness \cite{Kas03,Sch04}.

The simplest model taking local sources of current path deviations
into account is a current flowing along a narrow irregular path
below the atoms \footnote{This is equivalent to using the wire
edge model mentioned above but with a very small wire width and
equal current.}. Such a model gives reasonable agreement in the
slope of the PSD as $d$ is increased as can be expected as long as
$d$ is small compared to the relevant period $1/k$. Applying this
method over the full spectrum ($k>1/200$ $\mu$m$^{-1}$), we obtain
a local current flow fluctuation spectrum that scales as $\sim
1/k^2$. Microscopically well characterized wires will have to be
fabricated and tested to develop a more refined model explaining
the disorder potentials caused by local current deviations.

From our data and the simple local model we can estimate the rms
strength of the relative disorder potential and scale it to
different heights. At a surface distance of $d=10$ $\mu$m we find
the rms $\Delta B/B=3\times 10^{-5}$ ($<10^{-5}$) for spatial
frequencies $k>1/200$ $\mu$m$^{-1}$ ($k>1/50$ $\mu$m$^{-1}$). At
$d>30$ $\mu$m, where disorder potentials near electroplated wires
have been measured, $\Delta B/B$ is significantly smaller than the
measurement sensitivity in our case ($5\times 10^{-6}$). This
corresponds to a reduction by about two orders of magnitude.

To conclude, we have investigated magnetic disorder potentials
near lithographically fabricated current carrying wires using
quasi 1d BECs. The measured potentials are orders of magnitude
smaller than previously observed in other atom chip experiments,
which we attribute to our different chip fabrication method
resulting in much smoother and much more homogeneous wires. We
have strong evidence that the remaining potential roughness can be
attributed to deviations of the current flow from its nominal path
through the wire. We find that in addition to wire edge roughness
{\em local} fluctuations can be a dominating source of the
disorder potentials. Our method has a sensitivity for $\Delta B/B$
of better than $10^{-5}$ for a single point measurement,
corresponding to a deviation of the local current path smaller
than $10^{-5}$~rad. The strong scaling of the magnetic field
fluctuations with spatial frequency indicates a dominance of large
scale inhomogeneities which can be dealt with by improving the
fabrication. The smallness of high frequency fluctuations opens up
the way to $\mu$m scale quantum manipulation on atom chips.

We thank M. Brajdic and L. Della Pietra for help in the
experiments. This work was supported by the European Union,
contract numbers IST-2001-38863 (ACQP), HPRN-CT-2002-00304
(FASTNet), HPMF-CT-2002-02022, and HPRI-CT-1999-00114 (LSF) and
the Deutsche Forschungsgemeinschaft, contract number SCHM
1599/1-1.


\vspace{-0.5cm}

\begin{thebibliography}{31}
\expandafter\ifx\csname
natexlab\endcsname\relax\def\natexlab#1{#1}\fi
\expandafter\ifx\csname bibnamefont\endcsname\relax
  \def\bibnamefont#1{#1}\fi
\expandafter\ifx\csname bibfnamefont\endcsname\relax
  \def\bibfnamefont#1{#1}\fi
\expandafter\ifx\csname citenamefont\endcsname\relax
  \def\citenamefont#1{#1}\fi
\expandafter\ifx\csname url\endcsname\relax
  \def\url#1{\texttt{#1}}\fi
\expandafter\ifx\csname
urlprefix\endcsname\relax\def\urlprefix{URL }\fi
\providecommand{\bibinfo}[2]{#2}
\providecommand{\eprint}[2][]{\url{#2}}

\bibitem[{\citenamefont{Folman et~al.}(2002)\citenamefont{Folman, Kr{\"u}ger,
  Schmiedmayer, Denschlag, and Henkel}}]{Fol02}
\bibinfo{author}{\bibfnamefont{R.}~\bibnamefont{Folman}},
  \bibinfo{author}{\bibfnamefont{P.}~\bibnamefont{Kr{\"u}ger}},
  \bibinfo{author}{\bibfnamefont{J.}~\bibnamefont{Schmiedmayer}},
  \bibinfo{author}{\bibfnamefont{J.}~\bibnamefont{Denschlag}},
  \bibnamefont{and} \bibinfo{author}{\bibfnamefont{C.}~\bibnamefont{Henkel}},
  \bibinfo{journal}{Adv. At. Mol. Opt. Phys.} \textbf{\bibinfo{volume}{48}},
  \bibinfo{pages}{263} (\bibinfo{year}{2002}).

\bibitem[{\citenamefont{Reichel et~al.}(1999)\citenamefont{Reichel,
  H{{\"a}}nsel, and H{{\"a}}nsch}}]{Rei99}
\bibinfo{author}{\bibfnamefont{J.}~\bibnamefont{Reichel}},
  \bibinfo{author}{\bibfnamefont{W.}~\bibnamefont{H{{\"a}}nsel}},
  \bibnamefont{and} \bibinfo{author}{\bibfnamefont{T.~W.}
  \bibnamefont{H{{\"a}}nsch}}, \bibinfo{journal}{Phys. Rev. Lett.}
  \textbf{\bibinfo{volume}{83}}, \bibinfo{pages}{3398} (\bibinfo{year}{1999}).

\bibitem[{\citenamefont{H{\"a}nsel
  et~al.}(2001{\natexlab{a}})\citenamefont{H{\"a}nsel, Reichel, Hommelhoff, and
  H{\"a}nsch}}]{Hae01a}
\bibinfo{author}{\bibfnamefont{W.}~\bibnamefont{H{\"a}nsel}},
  \bibinfo{author}{\bibfnamefont{J.}~\bibnamefont{Reichel}},
  \bibinfo{author}{\bibfnamefont{P.}~\bibnamefont{Hommelhoff}},
  \bibnamefont{and} \bibinfo{author}{\bibfnamefont{T.~W.}
  \bibnamefont{H{\"a}nsch}}, \bibinfo{journal}{Phys. Rev. A}
  \textbf{\bibinfo{volume}{64}}, \bibinfo{pages}{063607}
  (\bibinfo{year}{2001}{\natexlab{a}}).

\bibitem[{\citenamefont{Reichel}(2002)}]{Rei02}
\bibinfo{author}{\bibfnamefont{J.}~\bibnamefont{Reichel}},
  \bibinfo{journal}{Appl. Phys. B} \textbf{\bibinfo{volume}{74}},
  \bibinfo{pages}{469} (\bibinfo{year}{2002}).

\bibitem[{\citenamefont{M{\"u}ller et~al.}(1999)\citenamefont{M{\"u}ller,
  Anderson, Grow, Schwindt, and Cornell}}]{Mue99}
\bibinfo{author}{\bibfnamefont{D.}~\bibnamefont{M{\"u}ller}},
  \bibinfo{author}{\bibfnamefont{D.~Z.} \bibnamefont{Anderson}},
  \bibinfo{author}{\bibfnamefont{R.~J.} \bibnamefont{Grow}},
  \bibinfo{author}{\bibfnamefont{P.~D.~D.} \bibnamefont{Schwindt}},
  \bibnamefont{and} \bibinfo{author}{\bibfnamefont{E.~A.}
  \bibnamefont{Cornell}}, \bibinfo{journal}{Phys. Rev. Lett.}
  \textbf{\bibinfo{volume}{83}}, \bibinfo{pages}{5194} (\bibinfo{year}{1999}).

\bibitem[{\citenamefont{Dekker et~al.}(2000)\citenamefont{Dekker, Lee, Lorent,
  Thywissen, Smith, Drndi\'c, Westervelt, and Prentiss}}]{Dek00}
\bibinfo{author}{\bibfnamefont{N.~H.} \bibnamefont{Dekker {\em et al.}}},
  \bibinfo{journal}{Phys. Rev. Lett.} \textbf{\bibinfo{volume}{84}},
  \bibinfo{pages}{1124} (\bibinfo{year}{2000}).

\bibitem[{\citenamefont{Folman et~al.}(2000)\citenamefont{Folman, Kr{\"u}ger,
  Cassettari, Hessmo, Maier, and Schmiedmayer}}]{Fol00}
\bibinfo{author}{\bibfnamefont{R.}~\bibnamefont{Folman}},
  \bibinfo{author}{\bibfnamefont{P.}~\bibnamefont{Kr{\"u}ger}},
  \bibinfo{author}{\bibfnamefont{D.}~\bibnamefont{Cassettari}},
  \bibinfo{author}{\bibfnamefont{B.}~\bibnamefont{Hessmo}},
  \bibinfo{author}{\bibfnamefont{T.}~\bibnamefont{Maier}}, \bibnamefont{and}
  \bibinfo{author}{\bibfnamefont{J.}~\bibnamefont{Schmiedmayer}},
  \bibinfo{journal}{Phys. Rev. Lett.} \textbf{\bibinfo{volume}{84}},
  \bibinfo{pages}{4749} (\bibinfo{year}{2000}).

\bibitem[{\citenamefont{Cassettari et~al.}(2000)\citenamefont{Cassettari,
  Hessmo, Folman, Maier, and Schmiedmayer}}]{Cas00}
\bibinfo{author}{\bibfnamefont{D.}~\bibnamefont{Cassettari}},
  \bibinfo{author}{\bibfnamefont{B.}~\bibnamefont{Hessmo}},
  \bibinfo{author}{\bibfnamefont{R.}~\bibnamefont{Folman}},
  \bibinfo{author}{\bibfnamefont{T.}~\bibnamefont{Maier}}, \bibnamefont{and}
  \bibinfo{author}{\bibfnamefont{J.}~\bibnamefont{Schmiedmayer}},
  \bibinfo{journal}{Phys. Rev. Lett.} \textbf{\bibinfo{volume}{85}},
  \bibinfo{pages}{5483} (\bibinfo{year}{2000}).

\bibitem[{\citenamefont{Kr{\"u}ger et~al.}(2003)\citenamefont{Kr{\"u}ger, Luo,
  Klein, Brugger, Haase, Wildermuth, Groth, Bar-Joseph, Folman, and
  Schmiedmayer}}]{Kru03}
\bibinfo{author}{\bibfnamefont{P.}~\bibnamefont{Kr{\"u}ger {\em et al.}}},
  \bibinfo{journal}{Phys. Rev. Lett.} \textbf{\bibinfo{volume}{91}},
  \bibinfo{pages}{233201} (\bibinfo{year}{2003}).

\bibitem[{\citenamefont{Treutlein et~al.}(2004)\citenamefont{Treutlein,
  Hommelhoff, Steinmetz, H{\"a}nsch, and Reichel}}]{Tre04}
\bibinfo{author}{\bibfnamefont{P.}~\bibnamefont{Treutlein}},
  \bibinfo{author}{\bibfnamefont{P.}~\bibnamefont{Hommelhoff}},
  \bibinfo{author}{\bibfnamefont{T.}~\bibnamefont{Steinmetz}},
  \bibinfo{author}{\bibfnamefont{T.~W.} \bibnamefont{H{\"a}nsch}},
  \bibnamefont{and} \bibinfo{author}{\bibfnamefont{J.}~\bibnamefont{Reichel}},
  \bibinfo{journal}{Phys. Rev. Lett.} \textbf{\bibinfo{volume}{92}},
  \bibinfo{pages}{203005} (\bibinfo{year}{2004}).

\bibitem[{\citenamefont{Ott et~al.}(2001)\citenamefont{Ott, Fortagh,
  Schlotterbeck, Grossmann, and Zimmermann}}]{Ott01}
\bibinfo{author}{\bibfnamefont{H.}~\bibnamefont{Ott}},
  \bibinfo{author}{\bibfnamefont{J.}~\bibnamefont{Fortagh}},
  \bibinfo{author}{\bibfnamefont{G.}~\bibnamefont{Schlotterbeck}},
  \bibinfo{author}{\bibfnamefont{A.}~\bibnamefont{Grossmann}},
  \bibnamefont{and}
  \bibinfo{author}{\bibfnamefont{C.}~\bibnamefont{Zimmermann}},
  \bibinfo{journal}{Phys. Rev. Lett.} \textbf{\bibinfo{volume}{87}},
  \bibinfo{pages}{230401} (\bibinfo{year}{2001}).

\bibitem[{\citenamefont{H{\"a}nsel
  et~al.}(2001{\natexlab{b}})\citenamefont{H{\"a}nsel, Hommelhoff, H{\"a}nsch,
  and Reichel}}]{Hae01b}
\bibinfo{author}{\bibfnamefont{W.}~\bibnamefont{H{\"a}nsel}},
  \bibinfo{author}{\bibfnamefont{P.}~\bibnamefont{Hommelhoff}},
  \bibinfo{author}{\bibfnamefont{T.~W.} \bibnamefont{H{\"a}nsch}},
  \bibnamefont{and} \bibinfo{author}{\bibfnamefont{J.}~\bibnamefont{Reichel}},
  \bibinfo{journal}{Nature} \textbf{\bibinfo{volume}{413}},
  \bibinfo{pages}{498} (\bibinfo{year}{2001}{\natexlab{b}}).

\bibitem[{\citenamefont{Leanhardt et~al.}(2002)\citenamefont{Leanhardt,
  Chikkatur, Kielpinski, Shin, Gustavson, Ketterle, and Pritchard}}]{Lea02}
\bibinfo{author}{\bibfnamefont{A.~E.} \bibnamefont{Leanhardt} {\em et al.}},
  \bibinfo{journal}{Phys. Rev. Lett.} \textbf{\bibinfo{volume}{89}},
  \bibinfo{pages}{040401} (\bibinfo{year}{2002}).

\bibitem[{\citenamefont{Schneider et~al.}(2003)\citenamefont{Schneider, Kasper,
  Hagen, Bartenstein, Engeser, Schumm, Bar-Joseph, Folman, Feenstra, and
  Schmiedmayer}}]{Sch03b}
\bibinfo{author}{\bibfnamefont{S.}~\bibnamefont{Schneider} {\em et al.}},
  \bibinfo{journal}{Phys. Rev. A} \textbf{\bibinfo{volume}{67}},
  \bibinfo{pages}{023612} (\bibinfo{year}{2003}).

\bibitem[{\citenamefont{Calarco et~al.}(2000)\citenamefont{Calarco, Hinds,
  Jaksch, Schmiedmayer, Cirac, and Zoller}}]{Cal00a}
\bibinfo{author}{\bibfnamefont{T.}~\bibnamefont{Calarco}},
  \bibinfo{author}{\bibfnamefont{E.~A.} \bibnamefont{Hinds}},
  \bibinfo{author}{\bibfnamefont{D.}~\bibnamefont{Jaksch}},
  \bibinfo{author}{\bibfnamefont{J.}~\bibnamefont{Schmiedmayer}},
  \bibinfo{author}{\bibfnamefont{J.~I.} \bibnamefont{Cirac}}, \bibnamefont{and}
  \bibinfo{author}{\bibfnamefont{P.}~\bibnamefont{Zoller}},
  \bibinfo{journal}{Phys. Rev. A} \textbf{\bibinfo{volume}{61}},
  \bibinfo{pages}{022304} (\bibinfo{year}{2000}).

\bibitem[{\citenamefont{Fortagh
  et~al.}(2002{\natexlab{a}})\citenamefont{Fortagh, Ott, Kraft, G{\"u}nther,
  and Zimmermann}}]{For02}
\bibinfo{author}{\bibfnamefont{J.}~\bibnamefont{Fortagh}},
  \bibinfo{author}{\bibfnamefont{H.}~\bibnamefont{Ott}},
  \bibinfo{author}{\bibfnamefont{S.}~\bibnamefont{Kraft}},
  \bibinfo{author}{\bibfnamefont{A.}~\bibnamefont{G{\"u}nther}},
  \bibnamefont{and}
  \bibinfo{author}{\bibfnamefont{C.}~\bibnamefont{Zimmermann}},
  \bibinfo{journal}{Phys. Rev. A} \textbf{\bibinfo{volume}{66}},
  \bibinfo{pages}{041604(R)} (\bibinfo{year}{2002}{\natexlab{a}}).

\bibitem[{\citenamefont{Leanhardt et~al.}(2003)\citenamefont{Leanhardt, Shin,
  Chikkatur, Kielpinski, Ketterle, and Pritchard}}]{Lea03}
\bibinfo{author}{\bibfnamefont{A.~E.} \bibnamefont{Leanhardt}},
  \bibinfo{author}{\bibfnamefont{Y.}~\bibnamefont{Shin}},
  \bibinfo{author}{\bibfnamefont{A.~P.} \bibnamefont{Chikkatur}},
  \bibinfo{author}{\bibfnamefont{D.}~\bibnamefont{Kielpinski}},
  \bibinfo{author}{\bibfnamefont{W.}~\bibnamefont{Ketterle}}, \bibnamefont{and}
  \bibinfo{author}{\bibfnamefont{D.~E.} \bibnamefont{Pritchard}},
  \bibinfo{journal}{Phys. Rev. Lett.} \textbf{\bibinfo{volume}{90}},
  \bibinfo{pages}{100404} (\bibinfo{year}{2003}).

\bibitem[{\citenamefont{Jones et~al.}(2003)\citenamefont{Jones, Vale, Sahagun,
  Hall, and Hinds}}]{Jon03}
\bibinfo{author}{\bibfnamefont{M.~P.~A.} \bibnamefont{Jones}},
  \bibinfo{author}{\bibfnamefont{C.~J.} \bibnamefont{Vale}},
  \bibinfo{author}{\bibfnamefont{D.}~\bibnamefont{Sahagun}},
  \bibinfo{author}{\bibfnamefont{B.~V.} \bibnamefont{Hall}}, \bibnamefont{and}
  \bibinfo{author}{\bibfnamefont{E.~A.} \bibnamefont{Hinds}},
  \bibinfo{journal}{Phys. Rev. Lett.} \textbf{\bibinfo{volume}{91}},
  \bibinfo{pages}{080401} (\bibinfo{year}{2003}).

\bibitem[{\citenamefont{Est{\`e}ve et~al.}(2004)\citenamefont{Est{\`e}ve,
  Aussibal, Schumm, Figl, Mailly, Bouchoule, Westbrook, and Aspect}}]{Est04}
\bibinfo{author}{\bibfnamefont{J.}~\bibnamefont{Est{\`e}ve} {\em et al.}},
  (\bibinfo{year}{2004}), \bibinfo{note}{physics/0403020}.

\bibitem[{\citenamefont{Drndi\'c et~al.}(1998)\citenamefont{Drndi\'c, Johnson,
  Thywissen, Prentiss, and Westervelt}}]{Drn98}
\bibinfo{author}{\bibfnamefont{M.}~\bibnamefont{Drndi\'c}},
  \bibinfo{author}{\bibfnamefont{K.~S.} \bibnamefont{Johnson}},
  \bibinfo{author}{\bibfnamefont{J.~H.} \bibnamefont{Thywissen}},
  \bibinfo{author}{\bibfnamefont{M.}~\bibnamefont{Prentiss}}, \bibnamefont{and}
  \bibinfo{author}{\bibfnamefont{R.~M.} \bibnamefont{Westervelt}},
  \bibinfo{journal}{Appl. Phys. Lett.} \textbf{\bibinfo{volume}{72}},
  \bibinfo{pages}{2906} (\bibinfo{year}{1998}).

\bibitem[{\citenamefont{Fortagh
  et~al.}(2002{\natexlab{b}})\citenamefont{Fortagh, Ott, Schlotterbeck,
  Zimmermann, Herzog, and Wharam}}]{For02b}
\bibinfo{author}{\bibfnamefont{J.}~\bibnamefont{Fortagh}},
  \bibinfo{author}{\bibfnamefont{H.}~\bibnamefont{Ott}},
  \bibinfo{author}{\bibfnamefont{G.}~\bibnamefont{Schlotterbeck}},
  \bibinfo{author}{\bibfnamefont{C.}~\bibnamefont{Zimmermann}},
  \bibinfo{author}{\bibfnamefont{B.}~\bibnamefont{Herzog}}, \bibnamefont{and}
  \bibinfo{author}{\bibfnamefont{D.}~\bibnamefont{Wharam}},
  \bibinfo{journal}{Appl. Phys. Lett.} \textbf{\bibinfo{volume}{81}},
  \bibinfo{pages}{1146} (\bibinfo{year}{2002}{\natexlab{b}}).

\bibitem[{\citenamefont{Lev}(2003)}]{Lev03}
\bibinfo{author}{\bibfnamefont{B.}~\bibnamefont{Lev}}, \bibinfo{journal}{Quant.
  Inf. Comp.} \textbf{\bibinfo{volume}{3}}, \bibinfo{pages}{450}
  (\bibinfo{year}{2003}).

\bibitem[{\citenamefont{Denschlag et~al.}(1999)\citenamefont{Denschlag,
  Cassettari, and Schmiedmayer}}]{Den99a}
\bibinfo{author}{\bibfnamefont{J.}~\bibnamefont{Denschlag}},
  \bibinfo{author}{\bibfnamefont{D.}~\bibnamefont{Cassettari}},
  \bibnamefont{and}
  \bibinfo{author}{\bibfnamefont{J.}~\bibnamefont{Schmiedmayer}},
  \bibinfo{journal}{Phys. Rev. Lett.} \textbf{\bibinfo{volume}{82}},
  \bibinfo{pages}{2014} (\bibinfo{year}{1999}).

\bibitem[{\citenamefont{Kasper et~al.}(2003)\citenamefont{Kasper, Schneider,
  vom Hagen, Bartenstein, Engeser, Schumm, Bar-Joseph, Folman, Feenstra, and
  Schmiedmayer}}]{Kas03}
\bibinfo{author}{\bibfnamefont{A.}~\bibnamefont{Kasper} {\em et al.}},
  \bibinfo{journal}{J. Opt. B} \textbf{\bibinfo{volume}{5}},
  \bibinfo{pages}{S143} (\bibinfo{year}{2003}).

\bibitem[{\citenamefont{Schumm et~al.}(2004)\citenamefont{Schumm, Est{\`e}ve,
  Figl, Trebbia, Aussibal, Nguyen, Mailly, Bouchoule, Westbrook, and
  Aspect}}]{Sch04}
\bibinfo{author}{\bibfnamefont{T.}~\bibnamefont{Schumm} {\em et al.}},
  (\bibinfo{year}{2004}), \bibinfo{note}{physics/0407094}.

\bibitem[{\citenamefont{Wang et~al.}(2004)\citenamefont{Wang, Lukin, and
  Demler}}]{Wan04}
\bibinfo{author}{\bibfnamefont{D.-W.} \bibnamefont{Wang}},
  \bibinfo{author}{\bibfnamefont{M.~D.} \bibnamefont{Lukin}}, \bibnamefont{and}
  \bibinfo{author}{\bibfnamefont{E.}~\bibnamefont{Demler}},
  \bibinfo{journal}{Phys. Rev. Lett.} \textbf{\bibinfo{volume}{92}},
  \bibinfo{eid}{076802} (\bibinfo{year}{2004}).

\bibitem[{\citenamefont{Groth et~al.}(2004)\citenamefont{Groth, Kr{\"u}ger,
  Wildermuth, Folman, Fernholz, Mahalu, Bar-Joseph, and Schmiedmayer}}]{Gro04}
\bibinfo{author}{\bibfnamefont{S.}~\bibnamefont{Groth} {\em et al.}},
  \bibinfo{journal}{Appl. Phys. Lett.}  (\bibinfo{year}{2004}),
  \bibinfo{note}{in press, cond-mat/0404141}.

\bibitem[{\citenamefont{Wildermuth et~al.}(2004)\citenamefont{Wildermuth,
  Kr{\"u}ger, Becker, Brajdic, Haupt, Kasper, Folman, and
  Schmiedmayer}}]{foot}
\bibinfo{author}{\bibnamefont{This
fabrication method \cite{Gro04} was used for the chip in all our
previous experiments \cite{Fol00,Cas00,Kru03}.}}

\bibitem[{\citenamefont{Wildermuth et~al.}(2004)\citenamefont{Wildermuth,
  Kr{\"u}ger, Becker, Brajdic, Haupt, Kasper, Folman, and
  Schmiedmayer}}]{Wil04}
\bibinfo{author}{\bibfnamefont{S.}~\bibnamefont{Wildermuth} {\em et al.}},
  \bibinfo{journal}{Phys. Rev. A} \textbf{\bibinfo{volume}{69}},
  \bibinfo{pages}{030901(R)} (\bibinfo{year}{2004}).

\bibitem[{\citenamefont{Bergeman et~al.}(2003)\citenamefont{Bergeman, Moore,
  and Olshanii}}]{Ber03}
\bibinfo{author}{\bibfnamefont{T.}~\bibnamefont{Bergeman}},
  \bibinfo{author}{\bibfnamefont{M.~G.} \bibnamefont{Moore}}, \bibnamefont{and}
  \bibinfo{author}{\bibfnamefont{M.}~\bibnamefont{Olshanii}},
  \bibinfo{journal}{Phys. Rev. Lett.} \textbf{\bibinfo{volume}{91}},
  \bibinfo{pages}{163201} (\bibinfo{year}{2003}).

\bibitem[{\citenamefont{Shin et~al.}(2004)\citenamefont{Shin, Saba, Schirotzek,
  Pasquini, Leanhardt, Pritchard, and Ketterle}}]{Shi04}
\bibinfo{author}{\bibfnamefont{Y.}~\bibnamefont{Shin} {\em et al.}},
  \bibinfo{journal}{Phys. Rev. Lett.} \textbf{\bibinfo{volume}{92}},
  \bibinfo{pages}{150401} (\bibinfo{year}{2004}).

\bibitem[{\citenamefont{McGuirk et~al.}(2004)\citenamefont{McGuirk, Harber,
  Obrecht, and Cornell}}]{McG04}
\bibinfo{author}{\bibfnamefont{J.~M.} \bibnamefont{McGuirk}},
  \bibinfo{author}{\bibfnamefont{D.~M.} \bibnamefont{Harber}},
  \bibinfo{author}{\bibfnamefont{J.~M.} \bibnamefont{Obrecht}},
  \bibnamefont{and} \bibinfo{author}{\bibfnamefont{E.~A.}
  \bibnamefont{Cornell}},
  \bibinfo{journal}{Phys. Rev. A} \textbf{\bibinfo{volume}{69}},
  \bibinfo{pages}{062905} (\bibinfo{year}{2004}).

\end{thebibliography}
\end{document}